\documentclass[a4paper,10pt]{article}

\usepackage[margin=1in]{geometry}

\pdfoutput=1

\usepackage{graphicx}
\usepackage{bm}
\usepackage{dcolumn}
\usepackage{setspace}
\usepackage{abstract}
\usepackage{multirow}
\usepackage{amsmath}
\usepackage{fancyhdr}

\usepackage[affil-it]{authblk}

\usepackage[utf8]{inputenc}
\usepackage{lineno,hyperref}

\hypersetup{colorlinks=true, linkcolor=blue, citecolor=red}

\begin{document}
\title{ \bf Molecular dynamics simulation of twin boundary effect
on deformation of Cu nanopillars}
\date{}

\author{\small G. Sainath\footnote{email : sg@igcar.gov.in}, B.K. Choudhary\footnote{email : sg@igcar.gov.in}}
\affil {Deformation and Damage Modelling Section, Mechanical Metallurgy 
Division \\ Indira Gandhi Centre for Atomic Research, Kalpakkam 603102, Tamilnadu, India}

\twocolumn[
\maketitle
\begin{onecolabstract}

Molecular dynamics simulations performed on $<$110$>$ Cu nanopillars revealed significant difference in 
deformation behavior of nanopillars with and without twin boundary. The plastic deformation in single 
crystal Cu nanopillar without twin boundary was dominated by twinning, whereas the introduction of twin 
boundary changed the deformation mode from twinning to slip consisting of leading partial followed by 
trailing partial dislocations. This difference in deformation behavior has been attributed to the 
formation of stair-rod dislocation and its dissociation in the twinned nanopillars.\\
 

\end{onecolabstract}
]
\renewcommand{\thefootnote}{\fnsymbol{footnote}} \footnotetext{* E-mail :
sg@igcar.gov.in}
\renewcommand{\thefootnote}{\fnsymbol{footnote}} \footnotetext{$\dagger$ E-mail :
bkc@igcar.gov.in}

{\small 

\section{Introduction}

The requirement to improve the mechanical properties of nanomaterials for advanced applications has 
raised renewed interest towards the twinned nanopillars. The twin boundaries (TBs) influence the 
strength as well as the deformation behavior of the nanopillars. It has been demonstrated that the 
twin boundary enhances the strength coupled with high ductility \cite{1}, improves crack resistance 
\cite{2} and increases strain rate sensitivity \cite{3}. The interaction of TBs with dislocations 
results in either strengthening or, softening or, has negligible effect on the strength value \cite{4,5}.
In addition to superior properties, the twinned nanopillars show novel deformation mechanisms \cite{6,7,8,9}. 
It has been shown that the 1/2$<$110$>$ dislocations glide on \{100\} plane instead of a conventional 
\{111\} plane after penetrating the twin boundary (TB) in Cu nanopillar \cite{6}. Based on the 
experimental results, Wang and Sui \cite{7} have shown that the leading and trailing partials can 
exchange their order after passing through the TB. TBs can also act as source/sink and glide plane 
for dislocations \cite{8}. Sinha and Kulkarni \cite{9} have demonstrated that the crack propagation 
along the twin boundary exhibits directional anisotropy in terms of cleavage in one direction and 
dislocation emission in the opposite direction. By performing in-situ tensile experiments, Jang et al. 
\cite{10} observed different mechanisms in Cu nanopillars with orthogonal and slanted twin boundaries. 
In nanopillar with orthogonal twin boundaries, twin boundary-dislocation interactions dominate plastic 
flow, whereas detwinning governs deformation in nanopillars with slanted twin boundaries \cite{10}.

Most of the studies reported in the literature have been performed on nanopillars containing orthogonal 
or, slanted twin boundaries. Using in-situ transmission electron microscopy, Roos et al. \cite{11} 
investigated the role of longitudinal twin boundary on the deformation mechanism of Au nanowires. Under 
tensile loading, they observed the storage of full dislocations in twinned nanowires, while twins and 
stacking faults were observed in single crystal nanowires. The transition in deformation mechanism from
twinning in single crystal to slip in the twinned nanowire has been attributed to the pile-up of leading 
partials against the twin boundary and nucleation of trailing partials on the same plane as that of the 
leading partials in the twinned nanowires \cite{11}. Motivated by this study, we performed molecular 
dynamics (MD) simulations to investigate the possibility of such transition in the deformation mechanism 
of $<$110$>$ Cu nanopillars with and without longitudinal TB. Based on the MD simulation results, we show
that the perfect $<$110$>$ Cu nanopillar deforms by twinning, while the introduction of longitudinal TB 
changes the deformation mode from twinning to full slip. Further, we explore the atomistic description of 
the novel deformation mechanism operating in the twinned nanopillars that is responsible for the observed 
transition.

\section{Simulation details}

MD simulations have been carried out in LAMMPS package \cite{12} using embedded atom method (EAM) potential 
for Cu given by Mishin et al. \cite{13}. The visualization was accomplished using AtomEye \cite{14}]. Burgers 
vector of dislocations were determined by dislocation extraction algorithm (DXA) \cite{15} and were assigned 
according to Thomson tetrahedron. Single crystal Cu nanopillars in oriented in $<$110$>$ axial direction with 
\{111\} and \{112\} as side surfaces were considered for this study. The twin boundary was introduced by $180^o$  
rotation about the [111] axis. The model nanopillars had a square cross section width (d) = 16 nm and consisted 
of about 673000 atoms. The pillar length (l) was twice the cross section width. No periodic boundary conditions 
were used in any direction. The model system was equilibrated to a temperature of 10 K in NVT ensemble. Upon 
completion of equilibrium process, the deformation was carried out in a displacement-controlled mode at a 
constant strain rate of $3.5\times10^8 s^{-1}$ by imposing displacements to atoms along the tensile axis that
varied linearly from zero at the bottom to a maximum value at the top layer. The average stress is calculated 
from the Virial expression \cite{16}.

\section{Results}

Fig. \ref{stress-strain} shows a stress–strain behavior of $<$110$>$ Cu nanopillars with and without longitudinal 
TB subjected to tensile loading. Both the nanopillars show similar stress-strain behavior during elastic deformation 
with an elastic modulus of 115 GPa. Following the elastic deformation up to peak stress, the yielding in nanopillars
leads to an abrupt drop in the flow stress. The yield stress of a Cu nanopillar with longitudinal TB is obtained as 
3 GPa, while that of perfect Cu nanopillar (without TB) the yield stress is observed as 2.6 GPa. Fig. \ref{stress-strain} 
also shows that the yield strength of Cu nanopillar containing a TB is marginally higher than the perfect nanopillar. 
These results indicate that the longitudinal twin boundaries increase the strength similar to that observed for nanopillar
with orthogonal twin boundaries \cite{1,17}. In orthogonally twinned nanopillars, the high yield strength is attributed 
to redistribution of interior stress \cite{1}, site specific dislocation nucleation \cite{4} and a strong repulsive 
force on the dislocations \cite{18}. However, in the longitudinally twinned nanopillars, the physical origin of high 
yield strength is not yet clear and this needs further investigation.

 \begin{figure}
  \centering
 \includegraphics[width=7cm]{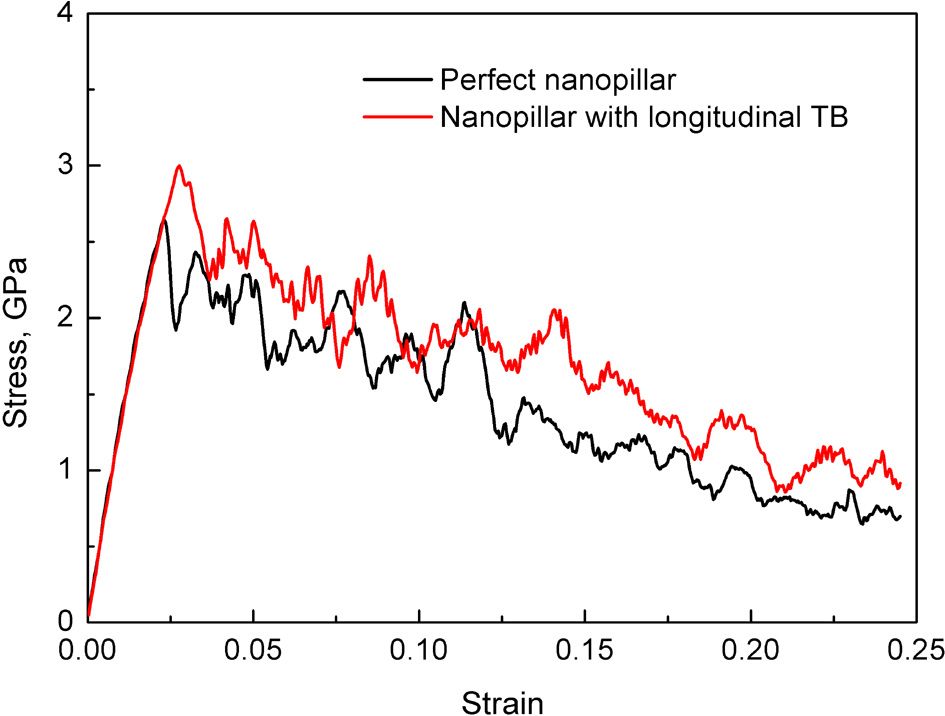}
 \caption {\footnotesize The stress-strain curves of $<$110$>$ axially oriented Cu nanopillars with and without longitudinal 
 twin boundary.}
 \label{stress-strain}
 \end{figure}
 
The atomic snapshots displaying the deformation behavior of perfect Cu nanopillar along with Thompson tetrahedron 
are shown in Fig. \ref{Pillar}. The Thompson tetrahedron (Fig. \ref{Pillar}(e)) is used to index the Burgers vectors 
and slip planes. It can be seen in Fig. \ref{Pillar}(a) that the nanopillar yields by the nucleation of four leading 
partials from the corners on two different \{111\} planes, $\alpha$ and $\beta$. Following the yielding, the stacking 
faults on planes $\alpha$ transform into twins (Fig. \ref{Pillar}(b)) by the successive nucleation of Shockley partials 
on adjacent $\alpha$ planes, while on planes $\beta$ they remains as stacking faults. With increasing deformation, one 
of the twins on planes $\alpha$ grows in width along with twin formation on a plane $\beta$ (Fig. \ref{Pillar}(c)). 
Further deformation is dominated entirely by the growth of twins on both $\alpha$ and $\beta$ planes as shown in 
Fig. \ref{Pillar}(d).

 \begin{figure}
  \centering
 \includegraphics[width=7.5cm]{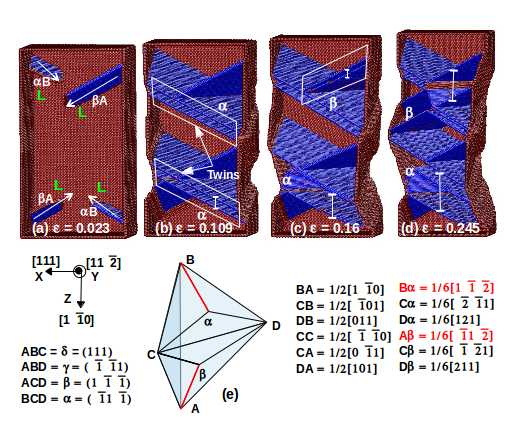}
 \caption {\footnotesize The deformation behavior of $<$110$>$ perfect Cu nanopillar along with Thompson tetrahedron. The 
 Thompson tetrahedron is orientated in such a way that the direction BA coincides with tensile axis and the plane ABC 
 is perpendicular to the [111] direction. The atoms are colored according to the common neighbor analysis \cite{19}. 
 The perfect fcc atoms are removed for clarity and atoms on stacking faults are shown in blue, while those of 
 dislocations and surface are shown in red. The alphabet L represents the leading partial. The white arrow represents 
 the direction of dislocation motion at that instant.}
 \label{Pillar}
 \end{figure}
 
The deformation behavior of a $<$110$>$ Cu nanopillar containing a longitudinal TB along with double Thompson tetrahedron is
shown in Fig. \ref{TB-Pillar}. The double Thompson tetrahedron (Fig. \ref{TB-Pillar}(f)) is used to index the Burgers vectors 
and slip planes activated during the deformation of a twinned nanopillar. As shown in Fig. \ref{TB-Pillar}(a), the twinned 
nanopillar yields by the nucleation of a leading partials with Burger vectors $\beta A$,  $A\beta'$ and $\alpha B$ from the 
corners on the planes $\beta$, $\beta'$ and $\alpha$, respectively. With increasing deformation, the leading partials 
$\beta A$ and $A\beta'$ meet at TB from the opposite side (Fig. \ref{TB-Pillar}(b)) and combine to form a stair-rod dislocation 
lying on TB with a Burgers vector $\beta\beta'$ as shown in inset in Fig. \ref{TB-Pillar}(b). It can be seen that the line of 
a stair-rod dislocation is parallel to CA, and its Burgers vector $\beta\beta'$ is perpendicular to both dislocation line and 
TB plane. Therefore, the stair-rod dislocation has an edge character having the magnitude $2a/3\sqrt{3}$ \cite{5}. Since the 
stair-rod dislocation is sessile on TB plane, with increasing deformation it dissociates into two trailing partials  $C\beta$ 
and $\beta'C$ as shown in Fig. \ref{TB-Pillar}(c). These trailing partials eliminate the stacking faults produced by leading 
partials on planes $\beta$ and $\beta'$. It can also be seen from Fig. \ref{TB-Pillar}(c) that the leading partial $B\alpha'$ 
also nucleates and interacts with $\alpha B$ at TB. This interaction leads to the formation of another stair-rod dislocation 
lying on TB with Burgers vector $\alpha\alpha'$. The line of this stair-rod dislocation is parallel to BC and is sessile on 
TB plane. Therefore, any further deformation leads to its dissociation into trailing partials, which eliminates the stacking 
fault on planes $\alpha$, $\alpha'$. The process of leading partial nucleation on symmetric slip planes, stair-rod formation 
followed by its dissociation into trailing partials is typically shown in Fig. \ref{Mechanism}. This process continues to 
occur at higher strains on different planes (Fig. \ref{TB-Pillar}(d)) and leads to the appearance of slip lines on surface 
of the twinned nanopillar (Fig. \ref{TB-Pillar}(e)). Due to the activation of symmetric slip systems, the formation of 
symmetrical slip lines about the TB can be seen in Fig. \ref{TB-Pillar}(e).

\begin{figure}
  \centering
 \includegraphics[width=7.5cm]{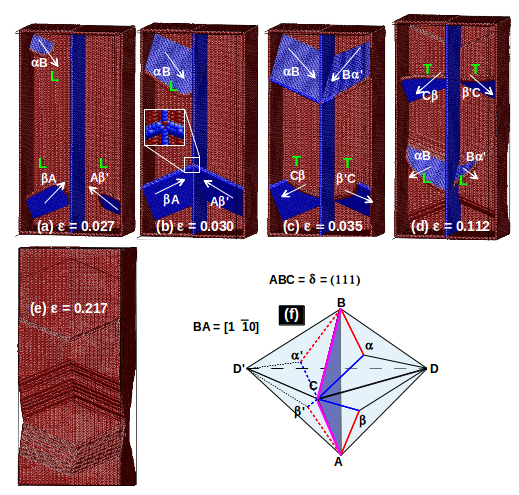}
 \caption {\footnotesize The deformation behavior of $<$110$>$ Cu nanopillar containing a longitudinal twin boundary along with 
 double Thompson tetrahedron. The double Thompson tetrahedron is orientated in such a way that, the direction BA coincides 
 with tensile axis and plane ABC is parallel to twin boundary plane. The alphabets L and T represents the leading and 
 trailing partials, respectively. The white arrow represents the direction of dislocation motion at that instant.}
 \label{TB-Pillar}
 \end{figure}
 
 \begin{figure}
  \centering
 \includegraphics[width=6cm]{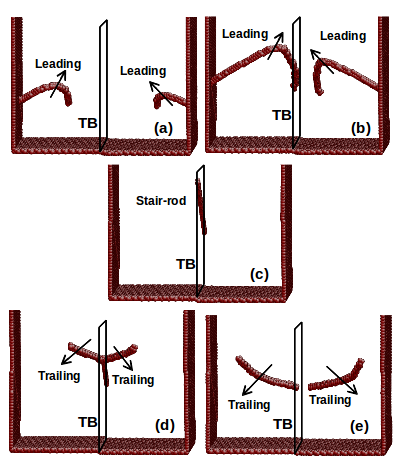}
 \caption {\footnotesize The process of stair-rod dislocation formation by leading partials and its dissociation into trailing partials. 
 The perfect fcc atoms and stacking faults (hcp atoms) are removed for clarity.}
 \label{Mechanism}
 \end{figure}

\section{Discussion}

The dominant mechanism of plastic deformation in perfect Cu nanopillar is through leading partial nucleation followed by 
twinning on two distinct twin systems [$\alpha$B]($\alpha$) and [$\beta$A]($\beta$). In agreement with our observations, 
many previous atomistic simulation studies \cite{20,21} have predicted twinning in $<$110$>$ oriented single crystal fcc 
metallic nanowires/nanopillars. Since the pillar axis coincides with line BA in Thompson tetrahedron (Fig. \ref{Pillar}(e)), 
the Schmid factor (m) is zero on $\gamma$ and $\delta$ planes and therefore these planes were not expected to participate 
in deformation. Moreover, the direction DC lying on the planes $\alpha$ and $\beta$ is perpendicular to the tensile axis 
and the slip will not occur along this direction also. The remaining possibilities are along BD and BC on plane $\alpha$
and AC and AD on plane $\beta$, i.e., a total of 4 systems. For $<$110$>$ tensile axis, the Schmid factors for leading 
($\alpha$B, $\beta$A) and trailing partial dislocations (C$\alpha$, $D\alpha$, $C\beta$, $D\beta$) are 0.471 and 0.235, 
respectively. Since, the Schmid factor for leading partial is higher than that of trailing partial, it is apparent that 
the nucleation of leading partials is easier than the trailing partials \cite{20}. Thus, the yielding in the perfect 
nanopillar occurs by the nucleation of leading partials alone. With increasing strain beyond yielding, the nucleation
of successive leading partials (can be called as twinning partials) on adjacent $\alpha$ and $\beta$ planes leads to 
deformation by twinning in $<$110$>$ oriented single crystal Cu nanopillar on two distinct twin systems 
[$\alpha$B]($\alpha$) and [$\beta$A]($\beta$).

The deformation in Cu nanopillar containing a longitudinal TB is dominated by slip involving both leading and trailing 
partial dislocations. It is important to mention that the MD simulations on $<$110$>$ Cu nanopillar containing two and 
more longitudinal TBs have shown similar deformation behavior dominated by slip. Further simulations and analysis are 
in progress. Interestingly, no twinning is observed in Cu nanopillar containing longitudinal TBs. The observation of 
deformation by slip in twinned Cu nanopillar and twinning in perfect Cu nanopillar is similar to experimental results 
on Au nanowires \cite{11}. In Au nanowires, the deformation by full slip in twinned nanopillar has been attributed to 
the pile-up of leading partials against the TB and nucleation of trailing partials on the same plane as that of the 
leading partials at higher stresses \cite{11}. Following an analogy with perfect nanopillar, the trailing partial
nucleation should not be observed in twinned nanopillar. However, due to the activation of symmetric slip systems, 
the two leading partials combine at TB to form a stair-rod dislocation, which dissociates into two trailing partials 
as observed in the present investigation in the twinned nanopillar (Fig. \ref{Mechanism}). In double Thompson tetrahedron, 
these processes can be written as

\begin{equation}
 \underbrace{\beta A}_\textit{leading} + \underbrace{A\beta'}_\textit{leading} \longrightarrow \underbrace{\beta\beta'}_\textit{stair-rod}
\end{equation}

The $\beta\beta'$ changes to $\beta'\beta$ due to twin symmetry (i.e $\beta\beta' = (\beta'\beta)_T$)

\begin{equation}
 \underbrace{\beta'\beta}_\textit{stair-rod} \longrightarrow \underbrace{C\beta}_\textit{trailing} + \underbrace{\beta'C}_\textit{trailing} 
\end{equation}

and similarly 

\begin{equation}
 \underbrace{\alpha B}_\textit{leading} + \underbrace{B\alpha'}_\textit{leading} \longrightarrow \underbrace{\alpha\alpha'}_\textit{stair-rod}
\end{equation}

\begin{equation}
 \underbrace{\alpha'\alpha}_\textit{stair-rod} \longrightarrow \underbrace{C\alpha}_\textit{trailing} + \underbrace{\alpha'C}_\textit{trailing} 
\end{equation}

The above process of stair-rod formation and its dissociation into trailing partials (Fig. \ref{Mechanism}) hinders the occurrence 
of twinning in twinned nanopillars and effectively results in full slip as described in the following reactions in the one half 
of nanopiller, 

\begin{equation}
 \underbrace{C\beta}_\textit{trailing} + \underbrace{\beta A}_\textit{leading} \longrightarrow  CA
 \end{equation}

\begin{equation}
 \underbrace{C\alpha}_\textit{trailing} + \underbrace{\alpha B}_\textit{leading} \longrightarrow  CB
\end{equation}

and in other half as 

\begin{equation}
 \underbrace{A\beta'}_\textit{leading} + \underbrace{\beta'C}_\textit{trailing} \longrightarrow  AC
\end{equation}

\begin{equation}
 \underbrace{B\alpha'}_\textit{leading} + \underbrace{\alpha'C}_\textit{trailing} \longrightarrow  BC
\end{equation}

The Burgers vector of a resultant full slip on $\beta$ and $\beta'$ planes (or $\alpha$ and $\alpha'$) are related 
by a twin or mirror symmetry ($CA = (AC)_{T}, CB = (BC)_{T}$) and they are parallel to TB that are shared by both 
the grains. This equivalence makes sure that both sides of the pillar deform collectively \cite{22}. In twinned Cu 
nanopillar, the slip occurs on four distinct slip systems, [CB]($\alpha$), [CA]($\beta$) and their symmetric 
counterparts [$BC$] ($\alpha'$), [$AC$]($\beta'$). The successive emission of leading partials followed by trailing 
partials and their escape at the free surface leads to the formation of well defined symmetrical slip steps with 
respect to the twin boundary on the surface of the twinned nanopillar. The observation of symmetrical slip lines is 
similar to that observed experimentally in micron sized Cu bicrystals \cite{22}.

\section{Conclusions}

Molecular dynamics simulations have been performed to understand the effect of longitudinal twin boundary on deformation
behavior of $<$110$>$ Cu nanopillar. The results showed that the single crystal nanopillar deforms by twinning on two 
independent twin systems, while the longitudinally twinned Cu nanopillar deforms by full slip with leading and trailing 
partial dislocations on four independent slip systems. The trailing partial nucleation in twinned nanopillar has been 
attributed to a novel mechanism of stair-rod formation by two leading partials followed by its dissociation. For
the first time using MD simulations, the atomistic description of the effect of longitudinal twin boundary on deformation 
behavior of $<$110$>$ Cu nanopillar has been provided.

}

\end{document}